\documentclass[11pt]{amsart}
\usepackage{braket}                
\usepackage{geometry}                
\usepackage{graphicx}
\usepackage{subcaption}
\usepackage{amssymb}
\usepackage{amsaddr}
\usepackage{epstopdf}
\usepackage{graphicx}
\title[Extending quantum theory with AS-assisted game theory]{Extending quantum theory with AI-assisted deterministic game theory (Extended abstract)}
\author{Florian Pauschitz (ETH Zurich), Ben Moseley (Imperial College), Ghislain Fourny (ETH Zurich)}

\begin{document}
\maketitle

We present an AI-assisted framework for predicting individual runs of quantum experiments, including contextuality and causality (adaptive), within our long-term programme of discovering a local hidden-variable theory that extends quantum theory.

Historically, the conclusions and interpretations drawn from impossibility theorems---such as Bell inequalities \cite{Bell1964}, the Kochen-Specker theorem \cite{Kochen1967}, the Free Will theorem \cite{Conway2006}, or constructs such as the Hardy model (possibilistic non-locality) \cite{Hardy1993} or the Greenberger–Horne–Zeilinger state (strong contextuality) \cite{Greenberger1989}--- have long been that local-realism must be abandoned, and that local hidden-variable models \cite{Nature2022} are inconsistent with what is observed in Nature. These conclusions rely on an often only implicitly mentioned assumption and remaining loophole: the free choice, by the observers, of the measurements to carry out. While some papers do mention it explicitly \cite{Renner2011}, it is at other times only found in the annex \cite{Storz2023}. Formally, this means that the choice of a measurement basis, modeled as a random variable located in spacetime, is statistically independent from any other random variables not situated in their future light cone. It is also otherwise known as the conjunction of measurement independence and parameter independence \cite{Sengupta2025}. Discussions of this assumption are often of a philosophical rather than of a mathematical or scientific nature, and often come down to an all-or-nothing argument---either free choice in this strong sense, or super-determinism \cite{Bell1980}\cite{Zeilinger2023}---and/or an emotional argument, e.g., ``In a world where everything could be explained by a conspiracy in the past, and where the experimenter would not be free to choose their parameters, I would give up the profession of physicist!'' (Alain Aspect \cite{Aspect2025}, translated from French; the exclamation mark is in the original sentence).

In previous work, we have shown that it is possible to frame the structure of complex quantum experiments (e.g., a process matrix \cite{Oreshkov2012} with fixed causal order together with some available choices of quantum instruments, or a causal contextuality scenario \cite{Abramsky2024}\cite{Fourny2025}) in terms of decision theory rather than probability theory \cite{Baczyk2024}. Formally, we designed a conversion algorithm from quantum experiment protocols into games in extensive form with imperfect information \cite{Kuhn1953}, a structure that is standard and mainstream in game theory and economics. Such games let observers play against the universe, which is seen as an economic agent minimizing action \cite{Fourny2020}. 

What sets our decision theory apart is that it is a deterministic local hidden-variable model; given the known rewards, the players act in a predictable way to optimally play the game. Thus, in order to obey the impossibility theorems and give quantum-consistent outcomes, this entails that we need to give up the notion of free choice as defined above. With our game-theoretical model, keeping free choice would mean resolving the game with a Nash resolution and a unilateral deviation assumption, which means no player can improve their reward by changing their strategy \emph{alone while all other players keep theirs fixed}. We showed that the resolution of such games with the classical Nash equilibrium framework does indeed check all the boxes of Bell inequalities: locality (via imperfect information), realism (via global mixed strategies), and free choice (via unilateral deviations). This means that if we are to use a framework to solve these game structures, it has to be non-Nashian \cite{Baczyk2024}. In other previous work, we designed non-Nashian solution concepts that drop the unilateral deviation assumption, which corresponds to free choice \cite{Fourny2018}\cite{Fourny2017}\cite{Fourny2019}. The opposite assumption is made: Perfect Prediction, meaning that the players predict each other successfully in all possible worlds; in modal logic, this means that the predictions are counterfactually dependent (not to be confused with retro-causality) on the decisions \cite{Dupuy2000}. This is a more subtle assumption than super-determinism, and it is linked with a compatibilistic form of free choice found in the philosophical literature \cite{Dupuy1992}. The resulting equilibrium, called the Perfectly Transparent Equilibrium (PTE), is at most unique for games in general position, and, when it exists, it is always Pareto-optimal.

With a game structure known for any fixed-causal-order experiment, and resolution algorithms in place, the last missing piece is deriving the reward functions, which associate every possible history with rewards for every player: observers, particles, etc, and determine the frequency (i.e., quantum probabilities) of the possible joint measurement outcomes conditioned on the joint settings. In this work, we focus on these reward functions in the particular context of the Einstein-Podolsky-Rosen (EPR) experiment \cite{Einstein1935}, as modified by Bohm to make it dichotomic \cite{Bohm2012}: 2 players, 2 choices of settings each, 2 outcomes possible. We study this experiment as it is a canonical test of non-locality and contextuality and will show, if we can determine suitable rewards, if our framework can obey impossibility theorems. We are currently not aware of an analytical approach for deriving reward functions, although there is ongoing research regarding the use of the principle of least action for spin measurements, e.g.,  \cite{Wen2023}. Instead, in this research, we lean on using AI to assist us to \textit{learn} rewards for a variety of quantum experiments. First, we represent the rewards with an unknown but learnable parameterized function (a neural network), which can optionally include any physics-inspired constraints we may choose to impose via its architecture design. Next, we build a differentiable PTE solver that, given the parameterized reward functions and a given quantum experiment in its game form, generates expected measurement settings and outcomes for many hidden variables samples. Finally, the parameters of our neural network reward function are learned with stochastic gradient descent by matching the observed outcome statistics, conditioned on settings, with expected statistics predicted by quantum theory. More precisely, we learn rewards that minimize the Kullback-Leibler divergence \cite{Kullback1951} between the histogram corresponding to the recorded frequencies of the outcomes (after millions or billions of deterministic runs with a random hidden variable), and the histogram corresponding to using the extended Born rule corresponding to the process matrix and the quantum instruments \cite{Born1926}\cite{Oreshkov2012}.

Several important features are required to ensure this algorithm converges stably and in a tractable fashion; we propose a more computationally efficient PTE solver, as well as a smooth relaxation of the solver by using annealed decision temperatures \cite{Pauschitz2024}, which allows for differentiability and training with backpropagation---which is the propagation of the difference between the two histograms all the way back to modified learned parameters in the reward functions, as is typical in gradient descent. We demonstrate that this approach exceeds classical physics and is able to closely reproduce quantum phenomena found in the EPR experiment. The framework successfully violates the Bell inequality despite the deterministic nature of the game solver. It is able to learn rewards that are closely consistent with expected quantum statistics and are physically meaningful. 

\begin{figure}[htbp]
    \centering
    \begin{subfigure}[b]{0.45\textwidth}
        \centering
\includegraphics[width=\textwidth]{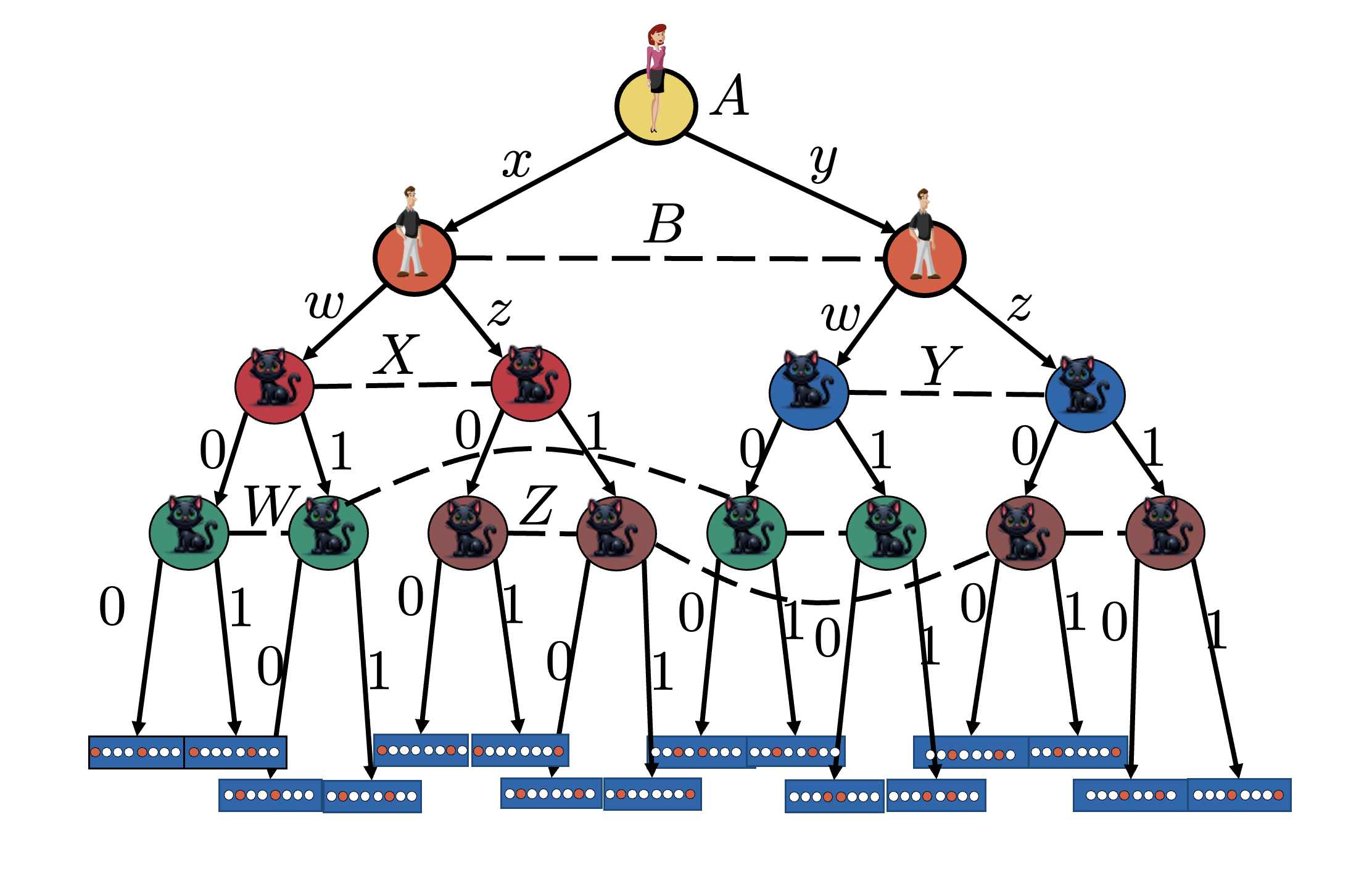}
        \caption{The game structure of the dichotomic version of the EPR experiment. It is a game in extensive form with imperfect information. Dashed lines (imperfect information) correspond, when interpreted in the Nash equilibrium framework, to locality and non-contextuality.}
        \label{fig:right}
    \end{subfigure}
    \hfill
    \begin{subfigure}[b]{0.45\textwidth}
        \centering
\includegraphics[width=\textwidth]{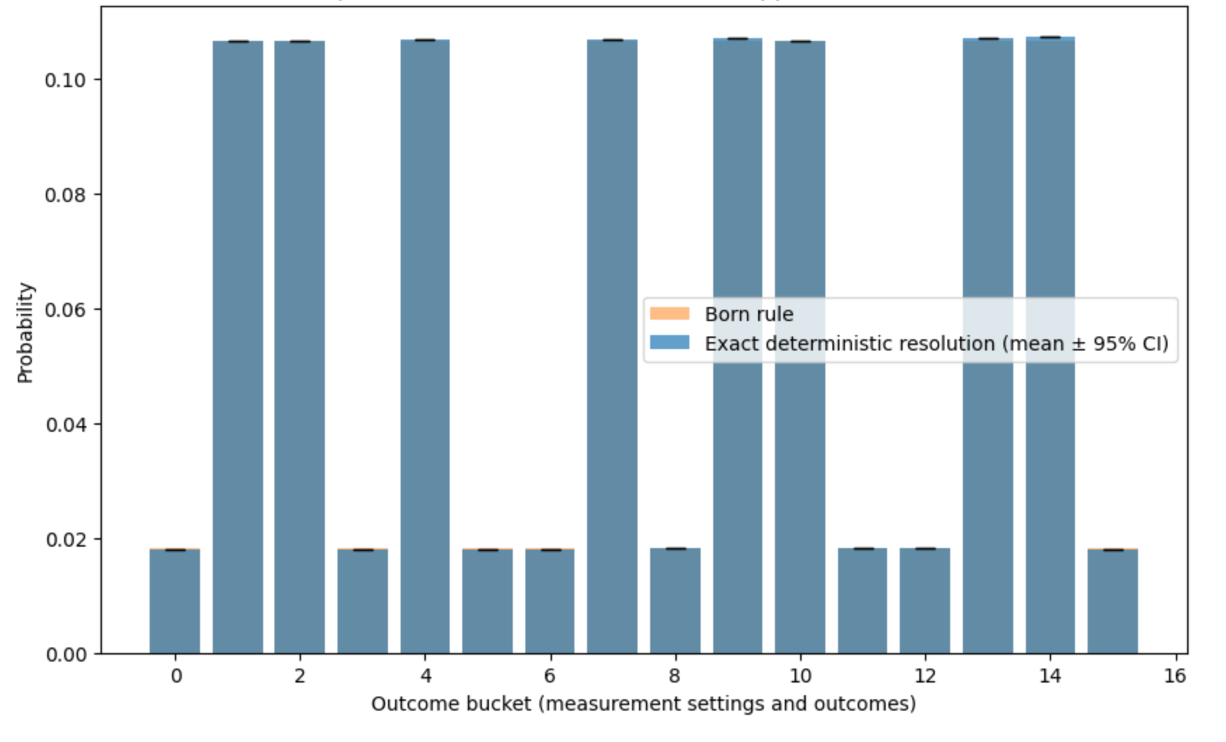}
        \caption{The closely coinciding histograms for the 2-2-2 EPR experiment, with the spin measurement angles taken with a 45-degree offset . The reason for the close, but not exact match is two fold: (i) because of the finite number of runs simulated, and (ii) because of the finite learning time with real parameter values.}
        \label{fig:left}
    \end{subfigure}
    \caption{The game structure and frequency histogram, which closely matches the Born rule.}
    \label{fig:both}
\end{figure}

While in this particular case, we obtained initial results by using a parameterized reward ansatz heavily inspired from Bell's original paper and were able to reduce the experimental space to a single dimension (the offset between the available choices of bases), we believe our underlying AI-assisted framework is generic enough to be used on more complex experiments where suitable ansatze do not exist, with more learnable parameters and involving alternate ways to learn (e.g., using symbolic regression, etc). Our results act as a proof-of-concept, and a toy local-realist hidden-variable model that non-Nashian quantum theory is a promising avenue towards a local hidden-variable theory. We discuss how analyzing the learned rewards for a larger, and more general set of quantum experiments, beyond the 2-2-2 EPR experiment, will help in inferring the local hidden-variable theory we are searching for. Our framework constitutes a solid foundation, which can be further expanded in order to fully discover a complete quantum theory.

\end{document}